\documentclass[fleqn]{article}% fleqn ensures left-alignment of equations
\usepackage[margin=1.2in]{geometry} %
\usepackage{mathtools}
\usepackage{amssymb}
\usepackage{newtxtext}
\usepackage{newtxmath}
\usepackage[table]{xcolor}
\usepackage{array}
\usepackage{diagbox}
\usepackage{booktabs}
\usepackage{mathrsfs}% e.g. nice Lagrange-L with \mathscr{L}
\usepackage{dsfont}% Blackboard letters

\usepackage{cite}%
\usepackage{nicefrac}% "nice" fraction
\usepackage{wrapfig}
\usepackage{enumitem}
\usepackage[pdftex,hidelinks]{hyperref}
\usepackage{pdfcomment}
\usepackage[tooltip]{acro}%\usepackage{acro}
\usepackage{xspace}
\usepackage[small]{caption}
\usepackage{subcaption}  % for \subcaptionbox
\usepackage{mleftright}
\mleftright
\usepackage[symbol]{footmisc}
\usepackage{microtype}
\usepackage{soul}

\newcommand*{\mytitle}{Modular Zeros}
\hypersetup{
    pdftitle = {\mytitle},
    pdfauthor = {Liu, Ratz}
}

\usepackage[noabbrev]{cleveref}% wants to be loaded last, use \Cref

\makeatletter
\g@addto@macro\bfseries{\boldmath}
\makeatother
\definecolor{c1}{RGB}{0,119,187}
\definecolor{c2}{RGB}{51,187,238}
\definecolor{c3}{RGB}{0,153,136}
\definecolor{c4}{RGB}{238,119,51}
\definecolor{c5}{RGB}{204,51,17}
\definecolor{c6}{RGB}{238,51,119}
\definecolor{c0}{RGB}{187,187,187}
\newcommand{\FilledEven}{\cellcolor{c4}}
\newcommand{\FilledOdd}{\cellcolor{c1}}

\newcommand{\ModulusT}{\ensuremath{\tau}}

\newcommand{\Generator}[1]{\ensuremath{\mathsf{#1}}}
\newcommand{\Group}[2]{\ensuremath{\text{#1}(#2)}}
\newcommand{\rep}[1]{\ensuremath{\boldsymbol{#1}}}

\DeclareMathOperator{\im}{Im}

\newcommand{\dd}{\mathop{}\!\mathrm{d}}
\newcommand{\ii}{\hskip0.1ex\mathrm{i}\hskip0.1ex}
\newcommand{\ee}{\mathrm{e}}% Euler e
\newcommand{\Id}{\ensuremath{\mathds{1}}}

\newcommand{\Z}[1]{\ensuremath{\mathds{Z}_{#1}}} % z_N ->\Z{N}

\makeatletter
\g@addto@macro\bfseries{\boldmath}% make section title math bold
\newcommand*{\defeq}{\mathchoice{\mathrel{\rlap{%
\raisebox{0.24ex}{$\m@th\cdot$}}%
\raisebox{-0.24ex}{$\m@th\cdot$}}%
=}{\mathrel{\rlap{%
\raisebox{0.24ex}{$\m@th\cdot$}}%
\raisebox{-0.24ex}{$\m@th\cdot$}}%
=}{\mathrel{\rlap{%
\raisebox{0.08ex}{\small$\m@th\cdot$}}%
\raisebox{-0.28ex}{\small$\m@th\cdot$}}%
=}{\mathrel{\rlap{%
\raisebox{0.08ex}{\tiny$\m@th\cdot$}}%
\raisebox{-0.28ex}{\tiny$\m@th\cdot$}}%
=}}
\newcommand*{\eqdef}{\mathchoice{=\mathrel{\rlap{%
\raisebox{0.24ex}{$\m@th\cdot$}}%
\raisebox{-0.24ex}{$\m@th\cdot$}}}{%
=\mathrel{\rlap{%
\raisebox{0.24ex}{$\m@th\cdot$}}%
\raisebox{-0.24ex}{$\m@th\cdot$}}}{%
=\mathrel{\rlap{%
\raisebox{0.08ex}{\small$\m@th\cdot$}}%
\raisebox{-0.28ex}{\small$\m@th\cdot$}}}{%
=\mathrel{\rlap{%
\raisebox{0.08ex}{\tiny$\m@th\cdot$}}%
\raisebox{-0.28ex}{\tiny$\m@th\cdot$}}}%
}
\newcommand*{\transpose}{%
{\mathpalette\@transpose{}}%
}
\newcommand*{\@transpose}[2]{%
\raisebox{\depth}{$\m@th#1\intercal$}%
}
\acsetup{pdfcomments/use=true}
\DeclareAcronym{BSM}{
  short = BSM ,
  long = beyond the standard model,
  pdfcomment = beyond the standard model
}
\DeclareAcronym{BU}{
  short = BU,
  long = bottom-up,
  pdfcomment = bottom-up
}
\DeclareAcronym{CFT}{
  short = CFT,
  long = conformal field theory,
  pdfcomment = conformal field theory
}
\DeclareAcronym{CG}{
  short = CG,
  long = Clebsch--Gordan,
  pdfcomment = Clebsch--Gordan
}
\DeclareAcronym{EFT}{
  short = EFT,
  long = effective field theory,
  pdfcomment = effective field theory
}
\DeclareAcronym{FI}{
  short = FI,
  long = Fayet--Iliopoulos \cite{Fayet:1974jb},
  pdfcomment = Fayet--Iliopoulos 
}
\DeclareAcronym{GS}{
  short = GS,
  long = Green--Schwarz \cite{Green:1984sg},
  pdfcomment = Green--Schwarz 
}
\DeclareAcronym{IR}{
  short = IR,
  long = infrared,
  pdfcomment = infrared
}
\DeclareAcronym{LEET}{
  short = LEET,
  long = low-energy effective theory,
  pdfcomment = low-energy effective theory
}
\DeclareAcronym{MSSM}{
  short = MSSM,
  long = minimal supersymmetric standard model,
  pdfcomment = minimal supersymmetric standard model
}
\DeclareAcronym{GUT}{
  short = GUT,
  long = grand unified theory,
  pdfcomment = grand unified theory
}
\DeclareAcronym{irrep}{
  short = irrep,
  long = irreducible representation,
  pdfcomment = irreducible representation
}
\DeclareAcronym{NNI}{
  short = NNI,
  long = nearest neighbor interaction,
  pdfcomment = nearest neighbor interaction
}
\DeclareAcronym{QFT}{
  short = QFT,
  long = quantum field theory,
  pdfcomment = quantum field theory
}
\DeclareAcronym{SM}{
  short = SM,
  long = standard model of particle physics,
  pdfcomment = standard model of particle physics
}
\DeclareAcronym{SUSY}{
  short = SUSY,
  long = supersymmetry,
  pdfcomment = supersymmetry
}
\DeclareAcronym{TD}{
  short = TD,
  long = top-down,
  pdfcomment = top-down
}
\DeclareAcronym{UV}{
  short = UV,
  long = ultraviolet,
  pdfcomment = ultraviolet
}
\DeclareAcronym{VEV}{
  short = VEV,
  long = vacuum expectation value,
  pdfcomment = vacuum expectation value
}
\DeclareAcronym{VVMF}{
  short = VVMF,
  long = vector-valued modular form,
  pdfcomment = vector-valued modular form
}

\definecolor{darkgreen}{HTML}{109930}
\definecolor{pink}{rgb}{0.858, 0.188, 0.478}

\title{\mytitle}
\numberwithin{equation}{section}
\numberwithin{figure}{section}
\numberwithin{table}{section}

\allowdisplaybreaks
\usepackage{braket}%Dirac Notation

\begin{document}

%%%%%%%%%%%%%%%%%%%%%%%%%%%%%%%%%%%%%%%%%%%%%%%%%%%%%%%%%%%%%%%%%%%%%%%%%%
%  Title page
%%%%%%%%%%%%%%%%%%%%%%%%%%%%%%%%%%%%%%%%%%%%%%%%%%%%%%%%%%%%%%%%%%%%%%%%%%
\begin{titlepage}
\vspace*{1.0cm}

\begin{flushright}
UCI--TR--2025--27%
\\
\end{flushright}

\vspace*{2cm}

\begin{center}
{\Huge\sffamily\bfseries\mytitle}

\vspace{1cm}

\renewcommand*{\thefootnote}{\fnsymbol{footnote}}

\textbf{%
Xiang-Gan Liu\footnote{xianggal@uci.edu}~~and~~Michael Ratz\footnote{mratz@uci.edu}
}
\\[8mm]
\textit{\small~Department of Physics and Astronomy, University of California, Irvine, CA 92697-4575 USA}
\end{center}

\vspace*{1cm}

\begin{abstract}
  Modular symmetries are known to be powerful and have various remarkable properties. 
  We point out that the structure of \acp{VVMF} space leads to the absence of couplings which cannot be explained in terms of the usual symmetries. 
  These modular zeros, which correspond to gaps in spaces of \acp{VVMF}, have the power of explaining certain stringy zeros, and to explain the renowned Weinberg texture that relates the Cabibbo angle to the hierarchies of the light down and strange quarks.
\end{abstract}

\vspace*{1cm}
\end{titlepage}
\renewcommand*{\thefootnote}{\arabic{footnote}}
\setcounter{footnote}{0}
\acresetall
%%%%%%%%%%%%%%%%%%%%%%%%%%%%%%%%%%%%%%%%%%%%%%%%%%%%%%%%%%%%%%%%%%%%%%%%%%
%  Intro
%%%%%%%%%%%%%%%%%%%%%%%%%%%%%%%%%%%%%%%%%%%%%%%%%%%%%%%%%%%%%%%%%%%%%%%%%%
\section{Introduction}
\label{sec:Introduction}

In modern physics, symmetry principles are central to determining, e.g.\ the structure of fundamental interactions and mass generation. 
It is well known that target space modular symmetries are a key ingredient of string theory \cite{Giveon:1994fu,DHoker:2022dxx}. 
Recently~\cite{Feruglio:2017spp}, modular symmetries have been used to explain the flavor structure of the \ac{SM} from a \ac{BU} approach; see e.g.\ \cite{Feruglio:2019ybq,Almumin:2022rml,Kobayashi:2023zzc,Ding:2023htn,Ding:2024ozt} for reviews. 
Congruence subgroups, which underlie modular flavor symmetries, emerge from orbifolding~\cite{Lauer:1989ax,Lerche:1989cs,Chun:1989se,Lauer:1990tm,Nilles:2020tdp,Nilles:2020gvu,Baur:2021mtl,Baur:2022hma,Li:2025bsr}. 
This leads to modular symmetries acting on the fields and Yukawa couplings in the \ac{EFT} description.
In particular, Yukawa couplings transform as \acp{VVMF}~\cite{Liu:2021gwa}. 
These \acp{VVMF} possess properties that have not yet been fully explored. 

The purpose of this Letter is to point out that gaps in the space of \acp{VVMF} allow us to understand the absence of terms in the Lagrange density that cannot be explained by usual symmetries.

The outline of the paper is as follows. After a short review of modular flavor symmetries in \Cref{sec:modular}, we introduce the concept of modular zeros in \Cref{sec:Modular_Zeros}. We discuss applications in \ac{TD} and \ac{BU} model building.
Finally, in \Cref{sec:conclusions} we provide a short summary of our results.

%%%%%%%%%%%%%%%%%%%%%%%%%%%%%%%%%%%%%%%%%%%%%%%%%%%%%%%%%%%%%%%%%%%%%%%%%%
\section{Modular Symmetries}
\label{sec:modular}

The (homogeneous) modular group $\Group{SL}{2,\mathds{Z}}$ can be defined via the presentation
\begin{equation}\label{eq:SL2Z}
\Gamma = \Group{SL}{2,\mathds{Z}} \defeq \Braket{\Generator{S},\Generator{T}\;|\;\Generator{S}^4 =(\Generator{S}\,\Generator{T})^3=\Id,\,\Generator{S}^2\,\Generator{T}=\Generator{T}\,\Generator{S}^2}\;,
\end{equation}
where the generators $\Generator{S}$ and $\Generator{T}$ can be represented as
\begin{equation}\label{eq:SandT}
\Generator{S} = \begin{pmatrix} 0 & 1 \\ -1 & 0 \end{pmatrix} \qquad\text{and}\qquad
\Generator{T} = \begin{pmatrix}
    1 & 1\\ 0 & 1
\end{pmatrix}\;.
\end{equation}
The modular group is accompanied by a modulus $\ModulusT$.
The modulus, the matter fields $\Phi$, and the superpotential $\mathscr{W}$ transform according to
\begin{subequations}\label{eq:ModularTransformationOfFieldsAndW}
\begin{align}
  \ModulusT &\xmapsto{~\gamma~} \frac{a\,\ModulusT + b}{c\,\ModulusT + d}\;, \label{eq:ModulusTrafo}\\
\Phi &\xmapsto{~\gamma~} (c\,\ModulusT + d)^{k_{\Phi}} \,\rho_{\Phi}(\gamma)\,\Phi\;,\label{eq:FieldTrafo} \\
\mathscr{W} &\xmapsto{~\gamma~}  (c\,\ModulusT + d)^{k_{\mathscr{W}}} \, \mathscr{W} \label{eq:superpotentialTrafo} \;,
\end{align} 
\end{subequations}
for  
\begin{equation}
  \gamma =\begin{pmatrix} a & b\\ c & d\end{pmatrix} \in \Group{SL}{2,\mathds{Z}}\;.
\end{equation}
Here $(c\,\ModulusT +d)^{k_{\Phi}}$ is the so-called automorphy factor, $k_{\mathscr{W}}\in\mathds{Z}$ is the modular weight of the superpotential, $k_{\Phi}\in\mathds{Q}$ is the modular weight of $\Phi$, and $\rho_{\Phi }(\gamma)$ is a matrix representation of the finite modular group $\Gamma_N$ or $\Gamma_N'$. 
In the supergravity context, the modular weight of the superpotential is generally nonzero. For convenience, we choose  $k_{\mathscr{W}}=-1$ throughout our work. For $2\leq N\leq5$, these finite modular groups can be defined by the presentations~\cite{Liu:2019khw}
\begin{subequations}\label{eq:ThetaTwistedMatter}
\begin{align}
\Gamma_N &\defeq \Braket{ \Generator{S},\Generator{T}\;|\;\Generator{S}^2 =(\Generator{S}\,\Generator{T})^3=\Generator{T}^N=\Id}\;, \label{eq:FiniteModularGroupsGammaN}\\
\Gamma_N'&  \defeq  \Braket{\Generator{S},\Generator{T}\;|\;\Generator{S}^4 =(\Generator{S}\,\Generator{T})^3=\Generator{T}^N=\Id,\;\Generator{S}^2\Generator{T}=\Generator{T}\,\Generator{S}^2}\;.\label{eq:FiniteModularGroupsGammaNprime}
\end{align}
\end{subequations}
\Cref{eq:ModularTransformationOfFieldsAndW} shows that the modulus $\ModulusT$ is sensitive only to $\Group{PSL}{2,\mathds{Z}}\cong \Group{SL}{2,\mathds{Z}} / \Z{2}$.
This is because $\ModulusT$ transforms identically under $\pm \gamma \in \Group{SL}{2, \mathds{Z}}$. 
Note, in particular, that $\ModulusT$ does not transform under $\Generator{S}^2=-\Id$.
In contrast, in general, the matter fields do distinguish between the transformations $\pm\gamma$.

The modular group $\Group{SL}{2,\mathds{Z}}$ allows one to define holomorphic functions known as \acp{VVMF}~\cite{Liu:2021gwa}, which depend on $\ModulusT$ in the upper half-plane $\mathcal{H}\defeq \left\{\ModulusT\in \mathds{C} \,|\, \im\ModulusT>0\right\}$, and are required to be regular at $\ii\infty$. 
Such \acp{VVMF} $\mathcal{Y}_{\rep{s}}^{(k_{\mathcal{Y}})}(\ModulusT)$ of modular weight $k_{\mathcal{Y}}\in\mathds{N}$ transform under $\gamma\in\Group{SL}{2,\mathds{Z}}$ according to
\begin{equation}
\label{eq:VVMFdef}
  \mathcal{Y}_{\rep{s}}^{(k_{\mathcal{Y}})}(\ModulusT) \xmapsto{~\gamma~} \mathcal{Y}_{\rep{s}}^{(k_{\mathcal{Y}})}(\gamma\,\ModulusT)= (c\,\ModulusT + d)^{k_{\mathcal{Y}}} \,\rho_{\rep{s}}(\gamma)\, \mathcal{Y}_{\rep{s}}^{(k_{\mathcal{Y}})}(\ModulusT)\;,
\end{equation}
where $\rho_{\rep{s}}(\gamma)$ is a unitary $s$-dimensional representation of a finite modular group, such as $\Gamma_N$ or $\Gamma_N'$. The \acp{VVMF} in the representation $\rho_{\rep{s}}$ and with weight $k_{\mathcal{Y}}$ form a vector space, denoted as $\mathcal{M}_{k_{\mathcal{Y}}}(\rho_{\rep{s}})$.
The direct sum $\mathcal{M}(\rho_{\rep{s}})=\bigoplus_{k_\mathcal{Y}\geq 0}\mathcal{M}_{k_{\mathcal{Y}}}(\rho_{\rep{s}})$ constitutes a graded free module. 
A crucial property of the module $\mathcal{M}(\rho_{\rep{s}})$ is that it is generated by the so-called minimal weight \ac{VVMF}, where the minimal weight $k^{(\rep{s})}_{\text{min}}$ is determined by the representation $\rho_{\rep{s}}$.
That is, all \acp{VVMF} of higher weight can be constructed from the minimal weight \ac{VVMF} through tensor products (an example of $T'$ can be found in  \Cref{app:Tprime}) or by the action of modular differential operators, etc. 
Therefore, the nonzero \acp{VVMF} in the representation $\rho_{\rep{s}}$ only exist when $k_{\mathcal{Y}} = k^{(\rep{s})}_{\text{min}} + 2n$ for $n\in \mathds{N}$. 
Additionally, for weights $k < 0$, there are no nonzero \acp{VVMF}, i.e.\ the space of \acp{VVMF} equals the empty set, $\mathcal{M}_{k_{\mathcal{Y}}<0}(\rho_{\rep{s}}) = \emptyset$. 
There are additional, less obvious, gaps for $k_{\mathcal{Y}}>0$ in the spaces of \acp{VVMF},
which will serve as the basis for the modular zeros discussed in this letter. 
Further properties of \ac{VVMF} have been discussed in \cite{Liu:2021gwa}.

\section{Modular Zeros: Gaps in the Modular Forms Space}
\label{sec:Modular_Zeros}

The absence of operators in Lagrange densities is usually a consequence of symmetries.
In many cases, this means that the corresponding term does not transform accordingly under a given symmetry operation. 

However, there are counter-examples to this statement. 
There are so-called stringy zeros \cite{Font:1988nc,Silverstein:1995re} which do not obey this pattern.
Furthermore, in supersymmetric theories texture zeros can actually be explained in terms of holomorphic zeros~\cite{Leurer:1993gy}. 
If one has a pseudo-anomalous $\Group{U}{1}_\mathrm{anom}$ symmetry, the \ac{FI} term must be cancelled by assigning \acp{VEV} to one or more fields with, say, positive $\Group{U}{1}_\mathrm{anom}$ charges $q_\mathrm{anom}$. 
Then, holomorphy and the sign of the charges $q_\mathrm{anom}$ lead to texture zeros. For more details, see \Cref{app:HolomorphicZeros}. 

In this section, we define a different type of texture zeros, which we refer to as \emph{modular zeros}. These arise from the underlying structure of \acp{VVMF}.

\subsection{Definition of Modular Zeros}
\label{subsec:Definition_of_Modular_Zeros}

Consider a superpotential of the form
\begin{equation}\label{eq:W_modular_zero}
  \mathscr{W} \supset Y^{(k_\mathcal{Y})}_{\rep{s}}(\ModulusT)\,\Phi_{I_1}\cdots\Phi_{I_n}\;.
\end{equation}
Modular invariance requires that the \acp{VVMF} $Y^{k_\mathcal{Y}}_{\rep{s}}(\ModulusT)$ belong to $\mathcal{M}_{k_\mathcal{Y}}(\rho_{\rep{s}})$ with $k_{\mathcal{Y}}+\sum_i k_{\Phi_{I_i}}=k_{\mathscr{W}}$ and $\rho_{\rep{s}} \otimes \rho_{\Phi_{I_1}}\otimes\dots \otimes \rho_{\Phi_{I_n}} \supset \rep{1}$.
We say that a \emph{modular zero} occurs if there is no appropriate \ac{VVMF} $Y^{(k_\mathcal{Y})}_{\rep{s}}(\ModulusT)$ that makes the term \eqref{eq:W_modular_zero} allowed, i.e.\ 
\begin{equation}\label{eq:ModularZero}
\mathcal{M}_{k_\mathcal{Y}}(\rho_{\rep{s}})=\emptyset\;.
\end{equation}
In this case, the coupling is absent. 
The vanishing is structural and follows from the absence of, or ``gap'' in, the space of 
\acp{VVMF} at the required weight $k_{\mathcal{Y}}$ and representation $\rho_{\rep{s}}$. 
Note that the modular zero applies to all orders of modulus $\ModulusT$ and is completely basis-independent.
Modular zeros occur, i.e.\ \Cref{eq:ModularZero} holds, in at least the following three situations:\footnote{For $k_{\mathcal{Y}}=0$, there exists only the constant modular form, which is in the trivial representation $\rho_{\rep{s}}=\rep{1}$.}
\begin{enumerate}[label={\Roman*.},ref={\Roman*}]
 \item $k_{\mathcal{Y}}<0$,\label{situation:1}
 \item $k_{\mathcal{Y}}> 0$, but \label{situation:2} 
  \begin{enumerate}[label={\Alph*.},ref={\Alph*}]
   \item $\rho_{\rep{s}}(\Generator{S}^2)\neq (-1)^{-k_{\mathcal{Y}}}\Id$, or\label{situation:2a}
   \item $k_{\mathcal{Y}}\neq k^{(\rep{s})}_{\mathrm{min}} +2 n$ for $n\in \mathds{N}$, \label{situation:2b}
  \end{enumerate} 
\end{enumerate}
where $k^{(\rep{s})}_{\text{min}}$ is the minimal weight for representation $\rho_{\rep{s}}$. 
The first situation parallels the above-mentioned holomorphic zeros.
On the other hand, the second situation, i.e.\ situation~\ref{situation:2}, relies on the intrinsic analytic properties of the modular forms and cannot be explained by a linear symmetry in the usual sense.
Situation \ref{situation:2}.\ref{situation:2a} will be discussed in \Cref{sec:Modular_spin}.
The case \ref{situation:2}.\ref{situation:2b} will be studied next, in an example.

\subsection{An Example}

In order to have a concrete example, consider a supersymmetric model based on the finite modular symmetry $\Gamma'_3\cong T'$ in which the \ac{MSSM} Higgs doublets, $H_u$ and $H_d$, have modular weight $k_H=-\nicefrac{5}{2}$. 
Assume first that both Higgs doublets furnish the representation $\rep{1'}$. 
Then, the $\mu$ term is given by\footnote{Note that the definition of modular zeros given in this subsection is also valid in the \ac{BU} framework, in which the superpotential typically carries zero modular weight. In the present example, the superpotential is a trivial singlet of $\Gamma'_3\cong T'$ and has modular weight $-1$.}
\begin{equation}\label{eq:mu_term_prime}
 \mathscr{W}_\mu = \lambda\,\bigl(\mathcal{Y}_{\rep{3}}^{(2)}\mathcal{Y}_{\rep{3}}^{(2)}\bigr)_{\rep{1'}}\,H_u\,H_d\;.
\end{equation}
Here, $\lambda$ is some normalization constant and the modular form $\mathcal{Y}_{\rep{3}}^{(2)}$ is given by the tensor product of the modular form doublet $\mathcal{Y}^{(1)}_{\rep{2''}}$ of weight 1
\begin{equation}\label{eq:Gamma_3_Y_3}
  \mathcal{Y}_{\rep{3}}^{(2)}\defeq\begin{pmatrix}
    \mathcal{Y}_{\rep{3},1}^{(2)}\\ \mathcal{Y}_{\rep{3},2}^{(2)}\\ \mathcal{Y}_{\rep{3},3}^{(2)}
  \end{pmatrix}= 
\begin{pmatrix}
-\sqrt{2}\, \bigl[\mathcal{Y}^{(1)}_{\rep{2''},1}\bigr]^{2} \\ 2\,  \mathcal{Y}^{(1)}_{\rep{2''},1}\,\mathcal{Y}^{(1)}_{\rep{2''},2} \\ \sqrt{2}\, \bigl[\mathcal{Y}^{(1)}_{\rep{2''},2}\bigr]^{2}
\end{pmatrix}\;,\quad\text{where}\quad \mathcal{Y}_{\rep{2''}}^{(1)}\defeq \begin{pmatrix}
\mathcal{Y}^{(1)}_{\rep{2''},1} \\
\mathcal{Y}^{(1)}_{\rep{2''},2}
\end{pmatrix}\;,
\end{equation}
see e.g.~\cite[Equation~(25)]{Chen:2024otk}. 
Furthermore, the contraction $\bigl(\mathcal{Y}_{\rep{3}}^{(2)}\mathcal{Y}_{\rep{3}}^{(2)}\bigr)_{\rep{1'}}$ is generically nontrivial. 
In fact, it yields a weight 4 modular form singlet $\mathcal{Y}_{\rep{1}'}^{(4)}=\eta^8(\tau)$. 
Thus, we obtain a nonzero bilinear form in \Cref{eq:mu_term_prime}.

However, if we assign the representation $\rep{1''}$ to $H_u$ and $H_d$, something interesting happens.
In this case, the bilinear emerges from
\begin{equation}\label{eq:mu_term_prime_prime}
  \mathscr{W}\supset \lambda\,\bigl(\mathcal{Y}_{\rep{3}}^{(2)}\mathcal{Y}_{\rep{3}}^{(2)}\bigr)_{\rep{1''}}\,H_u\,H_d\;.
\end{equation}
As already pointed out in \cite{Feruglio:2017spp}, the $\bigl(\mathcal{Y}_{\rep{3}}^{(2)}\mathcal{Y}_{\rep{3}}^{(2)}\bigr)_{\rep{1''}}$ contraction vanishes, i.e.\ 
\begin{equation}\label{eq:1_prime_prime}
  \bigl(\mathcal{Y}_{\rep{3}}^{(2)}\mathcal{Y}_{\rep{3}}^{(2)}\bigr)_{\rep{1''}}=
        \left(\mathcal{Y}_{\rep{3},2}^{(2)}\right)^2+2\mathcal{Y}_{\rep{3},1}^{(2)}\,\mathcal{Y}_{\rep{3},3}^{(2)}=0\;.
\end{equation}
This can be seen by inserting \Cref{eq:Gamma_3_Y_3} into \Cref{eq:1_prime_prime}. 

An appropriate $\mu$ term may be generated by the Kim--Nilles \cite{Kim:1983dt} and/or Giudice--Masiero mechanisms \cite{Giudice:1988yz}.
However, the absence of this coupling is not due to a symmetry in the usual sense. 
The \Z3 phase of the operator is trivial and, by the usual arguments, the coupling \Cref{eq:mu_term_prime_prime} should, therefore, be on the same footing as \Cref{eq:mu_term_prime}.
The vanishing of \Cref{eq:1_prime_prime} can instead be attributed to the fact that the triplet $\mathcal{Y}_{\rep{3}}^{(2)}$ in \Cref{eq:Gamma_3_Y_3} can be constructed from two independent modular forms $\mathcal{Y}^{(1)}_{\rep{2''},1}$ and $\mathcal{Y}^{(1)}_{\rep{2''},2}$~\cite{Liu:2019khw}, as shown in \Cref{eq:Gamma_3_Y_3}. 
As a result, the relations among modular forms are responsible for the vanishing of certain couplings and, hence, yield so-called modular zeros which cannot be explained by the conventional symmetries. 
In fact, this type of modular zero corresponds to situation \ref{situation:2}.\ref{situation:2b} described in the previous section. 
More specifically, from \Cref{tab:VVMFs} or \Cref{tab:GapsInTprime}, we can see that the minimal weight $k^{(\rep{1''})}_{\mathrm{min}}$ of the singlet modular form $\mathcal{Y}_{\rep{1''}}$ is 8, which is greater than the weight 4 of the Yukawa coupling in \Cref{eq:mu_term_prime_prime}. 
As a consequence, the coupling does not exist. 

Note that the modular zeros vary for each finite modular group, even if these finite modular groups are isomorphic in group structure. 
A specific example is the three different realizations of the modular $T'$~\cite{Arriaga-Osante:2025ppz}, whose \acsp{VVMF} spaces have very different ``gaps'', resulting in different texture zeros under the same charge assignment of matter fields; see \cite[Tables~5 and~6]{Arriaga-Osante:2025ppz}. 
In \Cref{sec:Modular_Texture_Zeros}, we will present a new application of modular zeros to texture zeros, which contrasts with the texture zeros achieved through traditional flavor symmetry realizations.

%%%%%%%%%%%%%%%%%%%%%%%%%%%%%%%%%%%%%%%%%%%%%%%%%%%%%%%%%%%%%%%%%%%%%%%%%%%%%%%%%%%%%%%%%%%%%%%%%%%%%%%%%%%%%%%%%%%%%
\subsection{Modular Spin}
\label{sec:Modular_spin}

Let us focus on the central element $\Generator{S}^2=-\Id$ of the modular group, which acts trivially on the modulus. 
Applying this central element in \Cref{eq:VVMFdef}, noting that the invariance of $\ModulusT$ leads to the invariance of the \acp{VVMF}, we have
\begin{equation}
\label{eq:Y-invarianceUnderS2-2}
  \mathcal{Y}_{\rep{s}}^{(k_{\mathcal{Y}})}(\ModulusT) \xmapsto{~\Generator{S}^2~}\mathcal{Y}_{\rep{s}}^{(k_{\mathcal{Y}})}(\ModulusT) = (-1)^{k_{\mathcal{Y}}} \rho_{\rep{s}}(\Generator{S}^2)\, \mathcal{Y}_{\rep{s}}^{(k_{\mathcal{Y}})}(\ModulusT)\;.
\end{equation}
This implies that the \Z2 subgroup generated by $\Generator{S}^2$, $\Z2^{\Generator{S}^2}$, imposes a constraint on any nonzero \acp{VVMF},
\begin{equation}
\label{eq:identityonYTrafo}
 (-1)^{k_{\mathcal{Y}}} \rho_{\rep{s}}(\Generator{S}^2) = \Id \qquad \iff\qquad
 \rho_{\rep{s}}(\Generator{S}^2) = (-1)^{-k_{\mathcal{Y}}}\, \Id\;.
\end{equation}
That is, the representation of \ac{VVMF} is closely related to its modular weight. 
As an example of this constraint, we can determine $\rho_{\rep{s}}(\Generator{S}^2)$ for the \acp{VVMF} of $\Gamma'_3\cong T'$. In \Cref{tab:VVMFs} we show these \acp{VVMF} for modular weights up to 9. We see that all of the even-dimensional representations $\rep{2}$, $\rep{2'}$ and  $\rep{2''}$ have odd weights, which implies $\rho_{\rep{s}}(\Generator{S}^2)=-\Id$. 
On the other hand, the odd-dimensional representations $\rep{1}$, $\rep{1'}$, $\rep{1''}$ and $\rep{3}$ come with even weights and, hence, $\rho_{\rep{s}}(\Generator{S}^2)=\Id$. 
This pattern is analogous to the statement that, in $\Group{SU}{2}$, odd powers of the fundamental spinor carry half-integer spin whereas even powers come with integer spin. 
The similarity of this scenario with spin transformations motivates the name \emph{modular spin} for the $\Z2^{\Generator{S}^2}$ symmetry.

Similarly, for matter fields $\Phi$ we see that
\begin{equation}
\label{eq:S2repMatter}
   \Phi \xmapsto{~\Generator{S}^2~} (-1)^{k_{\Phi}} \rho_{\Phi}(\Generator{S}^2)\, \Phi \;.
\end{equation}
Note that, unlike the case of the \acp{VVMF} (cf.\ \Cref{eq:Y-invarianceUnderS2-2}), $\rho_{\Phi}(\Generator{S}^2)$ is not fixed by the modular weight $k_{\Phi}$. 
However, since $\Generator{S}^2$ is always the \Z2 center of the finite modular group $\Gamma'_N$, its representation matrix $\rho_{\Phi} (\Generator{S}^2)$ can only take the values $\pm \Id$ for irreducible representations. 
Therefore, for integer weight $k_{\Phi}$, ${(-1)^{k_\Phi}}\,\rho_{\Phi} (\Generator{S}^2)$ can also only be $\pm \Id$. 
Yet \Cref{eq:S2repMatter} implies that, unlike the modulus $\tau$, matter fields can still transform nontrivially under $\Generator{S}^2$ if their modular weights are odd or fractional. 
As we shall see next, this has important consequences for discrete R-symmetries.

\subsection{Modular Spin and Discrete R-Symmetries}
\label{sec:discrete_R-symmetries}

In supergravity, the superpotential has nontrivial modular weight; modular symmetries are thus R-symmetries~\cite{Ibanez:1992uh}.
The transformation \eqref{eq:S2repMatter} distinguishes between the bosonic and fermionic components of chiral superfields. 
This provides us with additional motivation for the terminology ``Modular Spin''.
Under this transformation, the superpotential acquires a sign,
\begin{align}
   \mathscr{W}\xmapsto{~\Generator{S}^2~}(-1)^{k_{\mathscr{W}}}\,\mathscr{W}=-\mathscr{W}
\end{align}
since $k_{\mathscr{W}}=1$. The superspace coordinates $\theta$, therefore, pick up a factor of $\ii$,
\begin{align}
   \theta\xmapsto{~\Generator{S}^2~}(-1)^{k_{\theta}}\,\theta=\ii\theta
\end{align}
such that $\int\!\dd^2\theta\,\mathscr{W}$ is invariant. This fixes $k_{\theta}=\nicefrac{1}{2}$. Therefore, as far as $\theta$ and $\mathscr{W}$ are concerned, the $\Generator{S}^2$ transformation is a $\mathds{Z}_4^R$ symmetry. 
This symmetry gets enhanced in the presence of matter fields with fractional modular weights, cf.\ \Cref{eq:S2repMatter}. 

As a specific example, consider the $\mathds{T}^2/\mathds{Z}_3$ orbifold plane. 
The so-called complex structure modulus, $\ModulusT$, is fixed at $\omega$. 
There is a nontrivial stabilizer symmetry of $\ModulusT$, generated by $\Generator{S}\,\Generator{T}$, which has order 3, i.e.\ $(\Generator{S}\,\Generator{T})^3=\mathds{1}$. 
This is, again, an R-symmetry, under which the superpotential and superspace coordinates transform as 
\begin{subequations}
\begin{align}
  \mathscr{W}&\xmapsto{~\Generator{S}\,\Generator{T}~}\omega\,\mathscr{W}\;,\\
  \theta&\xmapsto{~\Generator{S}\,\Generator{T}~}\ee^{2\piup\ii/6}\theta\;. 
\end{align}  
\end{subequations} 
This means that $\Generator{S}\,\Generator{T}$ acts as a $\mathds{Z}_6^R$ on $\theta$.
Since the matter fields have fractional modular weights with denominator 3, this $\mathds{Z}_6^R$ gets enhanced to $\mathds{Z}_{18}^R$.\footnote{There are different conventions in the literature. For instance, in \cite{Baur:2024qzo} this symmetry is referred to as $\mathds{Z}_9^R$, which implies that $\theta$ carries a noninteger $\mathds{Z}_9^R$ charge of $\nicefrac{1}{2}$.}

Therefore, under $\Generator{S}^2\,(\Generator{S}\,\Generator{T})^{-1}$,
\begin{equation}
 \theta\xmapsto{~\Generator{S}^2\,(\Generator{S}\,\Generator{T})^{-1}~}\ee^{2\piup\ii/12}\,\theta\;, 
\end{equation}
so $\theta$ sees a $\mathds{Z}_{12}^R$ symmetry. 
The fractional weights of the matter fields then enhance this to a $\mathds{Z}_{36}^R$ symmetry\rlap{,}\footnote{In the conventions of \cite{Baur:2024qzo}, this is a $\mathds{Z}_{18}^R$ symmetry, again with $\theta$ carrying charge $\nicefrac{1}{2}$} a fact that has, to our knowledge, not been discussed in the literature.

An interesting aspect of this analysis is that the R-charges of the fields can be obtained from their modular weights and representations under the finite modular group, $\rho_\Phi(\gamma)$, e.g.\ $\rho_\Phi\bigl(\Generator{S}^2\,(\Generator{S}\,\Generator{T})^{2}\bigr)$ in the above example of the $\Z3$ orbifold. 
This approach is somewhat complementary to the traditional methods~\cite{Dixon:1986qv,Hamidi:1986vh,CaboBizet:2013gns,Nilles:2013lda}, and will be carried out elsewhere.

\subsection{Stringy Zeros}
\label{sec:Stringy_Zeros}

In string theory, certain couplings are absent but appear not to be forbidden by known symmetries~\cite{Font:1988nc,Silverstein:1995re,Anderson:2021unr,Gray:2024xun}. 
Specifically, couplings in orbifold compactifications must satisfy the \ac{CFT} selection rule known as ``Rule 4''~\cite{Font:1988nc,Kobayashi:2011cw}. 
Rule 4 dictates that local correlation functions vanish when all vertex operators are located at the same fixed point, depending on the presence of oscillator modes and derivatives~\cite{Font:1988nc}. 
Previous work has shown that neither continuous symmetries, discrete non-R-symmetries, nor Abelian discrete R-symmetries can explain this restriction~\cite{Font:1988nc}. 

\begin{wrapfigure}[15]{r}{6cm}
\centering\includegraphics{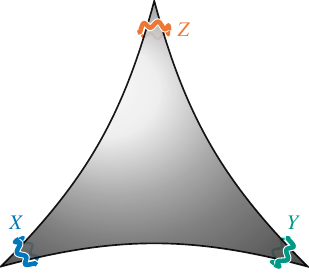}
\caption{Cartoon of the $\mathds{T}^2/\Z3$ orbifold pillow.}
\label{fig:Z3_orbifold_ravioli}
\end{wrapfigure}

We examine the \texorpdfstring{$\mathds{T}^2/\Z3$}{T**2/Z3} orbifold plane to illustrate this. 
Its three fixed points carry twisted states and can be visualized as the corners of a ``pillow'' (see \Cref{fig:Z3_orbifold_ravioli}). 
In the ``localization basis,'' twisted sector states are localized at their respective fixed points~\cite{Chun:1989se,Li:2025bsr}, and we refer to these fields as ``localization eigenstates.''
We consider the twisted field $\Phi_{-\nicefrac{5}{3}}^{(i)}\defeq \bigl( \widetilde{X}, \widetilde{Y}, \widetilde{Z} \bigr)^{\transpose}$ (with an oscillator mode) and $\Phi_{-\nicefrac{2}{3}}^{(i)} \defeq \bigl(X, Y, Z \bigr)^{\transpose}$ (without), where the superscript $(i=1\dots3)$ denotes the triplet component.

In the \texorpdfstring{$\mathds{T}^2/\Z3$}{T**2/Z3} orbifold, the picture changing mechanism for $N$-point Yukawa couplings yields $N-3$ derivatives. 
Since fields with oscillator modes carry additional derivatives, Rule 4 constrains the total derivative sum $\mathnormal{\Sigma}$.
Specifically, the rule states:
\begin{quote}
 \hypertarget{Rule4}{\textbf{Rule 4}:} \em Couplings of twisted fields at the same fixed point are only allowed if the sum of the number of derivatives $\mathnormal{\Sigma}$ equals $0 \bmod 6$.
\end{quote}
\mbox{}We can compute $\mathnormal{\Sigma}$ explicitly. 
Each oscillator mode contributes one derivative; thus, the monomial $\bigl(\Phi_{-\nicefrac{5}{3}}^{(i)}\bigr)^N$ adds $N$ derivatives. 
The total sum is
\begin{align}
\MoveEqLeft
    \mathnormal{\Sigma} = \begin{dcases}
        N-3&\text{for couplings involving only non-oscillator states with }k_\Phi=-\nicefrac{2}{3}\;,\\
     2N -3&\text{for couplings involving only oscillator states with }k_\Phi=-\nicefrac{5}{3}\;.
    \end{dcases}
    &\label{eq:Sigma}
\end{align}
Consequently, \hyperlink{Rule4}{Rule 4} imposes $\mathnormal{\Sigma}\overset{!}{=}0\bmod6$.
For oscillator modes, $\mathnormal{\Sigma}=2N-3$ is always odd, meaning \hyperlink{Rule4}{Rule 4} forbids all local oscillator couplings. 
For non-oscillator modes, the constraint implies $N=3\bmod6$, allowing only monomials with powers that are odd multiples of $3$. 
In what follows, we will argue that the modular spin provides the missing explanation.

Since the component labeling of $\Phi_{k_{\Phi}}^{(i)}$ is basis-dependent, we assume no loss of generality by studying $\bigl(\Phi_{-\nicefrac{2}{3}}^{(1)}\bigr)^N = X^N$ and $\bigl(\Phi_{-\nicefrac{5}{3}}^{(1)}\bigr)^N = \widetilde{X}^N$.
While a $\mathds{Z}_3$ permutation symmetry exists in the absence of Wilson lines~\cite{Kobayashi:2006wq,Nilles:2020tdp,Baur:2022hma,Li:2025bsr}, our conclusions hold regardless of this symmetry, relying only on the freedom to choose the $\mathds{Z}_3$ lattice basis. 
To form an allowed coupling, we multiply the monomial $\bigl(\Phi_{k_{\Phi}}^{(1)}\bigr)^N$ by a \ac{VVMF} $\mathcal{Y}_{\rep{s}}^{(k_{\mathcal{Y}})}(\ModulusT)$. 
Under $\Generator{S}^2$, the term transforms as
\begin{align}
\mathcal{Y}_{\rep{s}}^{(k_{\mathcal{Y}})}(\ModulusT) \,  \bigl(\Phi_{k_{\Phi}}^{(1)}\bigr)^N &\xmapsto{~\Generator{S}^2~} \bigl( -1 \bigr)^{k_{\mathcal{Y}}} \, \rho_{\rep{s}} \bigl( \Generator{S}^2 \bigr) \, \mathcal{Y}_{\rep{s}}^{(k_{\mathcal{Y}})}(\ModulusT) \, (-1)^{N k_{\Phi}} \, \rho_{\Phi}^{N}\bigl( \Generator{S}^2 \bigr) \,  \bigl(\Phi_{k_{\Phi}}^{(1)}\bigr)^N \nonumber \\
 &\overset{!}{=}\bigl(-1 \bigr)^{-1}\mathcal{Y}_{\rep{s}}^{(k_{\mathcal{Y}})}(\ModulusT) \,  \bigl(\Phi_{k_{\Phi}}^{(1)}\bigr)^N\;,
\label{eq:FullTrafo}
\end{align}
where we used \Cref{eq:Y-invarianceUnderS2-2,eq:S2repMatter,eq:superpotentialTrafo}. 
Matching the modular weights to the superpotential weight ($-1$) yields
\begin{equation}
   k_{\mathcal{Y}}+N\, k_{\Phi}\overset{!}{=}-1\;.
   \label{eq:modWightsCondition}
\end{equation}
Since the modular forms of $\Gamma_3'$ have positive integer modular weights $k_{\mathcal{Y}}$, this condition leads to a constraint on $N$. 
Substituting $k_{\Phi} = -\nicefrac{2}{3}$ or $-\nicefrac{5}{3}$ requires $N=3n$ for some positive integer $n$, leading to
\begin{equation}
\label{eq:kY}
  k_{\mathcal{Y}}
  =\begin{dcases}
    2 n -1 &\text{for couplings involving only non-oscillator states with }k_\Phi=-\nicefrac{2}{3}\;,\\
    5n-1&\text{for couplings involving only oscillator states with }k_\Phi=-\nicefrac{5}{3}\;.
  \end{dcases}
\end{equation}
Additionally, the product of representation matrices $\rho_{\Phi}^{3n}\bigl(\Generator{S}^2 \bigr)$ and $\rho_{\rep{s}}\bigl( \Generator{S}^2 \bigr)$ must leave the coupling in \Cref{eq:FullTrafo} invariant.
The modular form matrix $\rho_s \bigl( S^2 \bigr)$ is given by \Cref{eq:identityonYTrafo}. 
The twisted field transformation is given by
\begin{equation}
\label{eq:rhoSTwistedComponent}
\rho_{\Phi}\bigl(\Generator{S}^2\bigr) = -1\;.
\end{equation}
Enforcing trivial action for the product yields
\begin{equation}
\label{eq:S2inv}
  \rho_{ \Phi }^{3n} \bigl( \Generator{S}^2 \bigr)\, \rho_s \bigl( \Generator{S}^2 \bigr) = (-1)^{3n} \, (-1)^{-k_\mathcal{Y}} \overset{!}{=} 1 \;.
\end{equation}
Substituting $k_{\mathcal{Y}}$ from \Cref{eq:kY} into \Cref{eq:S2inv} gives 
\begin{equation}\label{eq:PowersM}
  1 \overset{!}{=} \begin{dcases}
    (-1)^{n+1}&\text{for couplings involving only non-oscillator states with }k_\Phi=-\nicefrac{2}{3}\;,\\
    (-1)^{-2 n+ 1}&\text{for couplings involving only oscillator states with }k_\Phi=-\nicefrac{5}{3}\;.
  \end{dcases}
\end{equation}
For oscillator states, $-2n+1$ is odd, prohibiting local couplings. 
For non-oscillator states, even powers are prohibited, allowing only odd multiples of $3$. 
Thus, the modular spin analysis reproduces the structure of \hyperlink{Rule4}{Rule 4}.
While \hyperlink{Rule4}{Rule 4} was previously thought to defy explanation by ordinary symmetries, we have shown that, at least for the $\Z3$ orbifold, it arises naturally as a modular zero: for the terms forbidden by \hyperlink{Rule4}{Rule 4} there is simply no appropriate \ac{VVMF} that can multiply the monomial of superfields. That is, stringy zeros do in fact have an explanation in terms of modular zeros.

A comprehensive analysis of different orbifold geometries with and without background fields as well as complete 6d examples is left for future work. 
Moreover, in the more general compactification of heterotic string, stringy zeros have been extensively studied in the form of vanishing theorems for Yukawa couplings~\cite{Anderson:2021unr,Gray:2024xun}. 
It will be interesting to investigate the possible connections between these vanishing theorems and the modular zeros discussed in this letter.

%%%%%%%%%%%%%%%%%%%%%%%%%%%%%%%%%%%%%%%%%%%%%%%%%%%%%%%%%%%%%%%%%%%%%%%%%%
%  Applications
%%%%%%%%%%%%%%%%%%%%%%%%%%%%%%%%%%%%%%%%%%%%%%%%%%%%%%%%%%%%%%%%%%%%%%%%%%

\subsection{Modular Texture Zeros}
\label{sec:Modular_Texture_Zeros}

Another interesting application of modular zeros concerns flavor physics.
Specifically, texture zeros are widely used in flavor model building. 
They can also be utilized to address the strong $\mathcal{CP}$ problem~\cite{Nelson:1983zb, Barr:1984qx, Barr:1984fh, Dine:2015jga}. 
A simple example was proposed by Weinberg~\cite{Weinberg:1977hb}. 
The aim is to explain a mass matrix 
\begin{equation}\label{eq:toy_mass_matrix}
 M_\text{Weinberg}=\begin{pmatrix}
  0 & \delta\\ 
  \delta & \mu 
 \end{pmatrix}\;,
\end{equation}
between, for example, quarks, i.e.\ 
\begin{equation}
  \mathscr{L}\supset -q_\mathrm{L}^i\, M_\text{Weinberg}^{ij}\,\bar d_\mathrm{R}^j\;.  
\end{equation}
In \Cref{eq:toy_mass_matrix}, we consider $\mu \gg \delta $. 
The texture zero in the $1$-$1$ entry is hard to explain with conventional symmetries. Following the holomorphic zeros approach (see \Cref{app:HolomorphicZeros}), the mass texture in \Cref{eq:toy_mass_matrix} can be obtained through a pseudo-anomalous $\Group{U}{1}_\mathrm{anom}$ symmetry. 
If we assign the quarks and the flavon $\phi$ the $\Group{U}{1}_\mathrm{anom}$ charges
\begin{equation}\label{eq:toy_example_holomorphic_zeros}
  q_{Q_\mathrm{L}}=\begin{pmatrix}
    1\\ -2   
  \end{pmatrix}\;,\quad
  q_{\bar d_\mathrm{R}}=\begin{pmatrix}
    1\\ -2   
  \end{pmatrix}\quad\text{and}\quad q_\phi=1\;,
\end{equation}
then the mass matrix takes the form 
\begin{equation}\label{eq:2x2FNmass}
 M_\text{holomorphic zero}=\begin{pmatrix}
    0 & \varepsilon\\ 
    \varepsilon & \varepsilon^4 
   \end{pmatrix}\;.
\end{equation}
Here $\varepsilon\defeq\Braket{\phi}/\Lambda$, where $\Lambda$ is some \ac{UV} scale, and we have ignored order-one coefficients.  
Clearly, the $1$-$1$ entry is forbidden while the other entries are allowed. 
However, the $2$-$2$ entry is suppressed relative to the off-diagonal entries, whereas Weinberg's texture \eqref{eq:toy_mass_matrix} requires the opposite hierarchy. 

This explanation is similar to the more recent one using the sign of modular weights~\cite{Feruglio:2023uof,Petcov:2024vph,Penedo:2024gtb,Feruglio:2024ytl}.
It suffices to label the charges of the quarks in \Cref{eq:toy_example_holomorphic_zeros} as modular weights. 
Since the modular forms have positive weights, the $1$-$1$ entry is again forbidden while the other entries are allowed.\footnote{This type of texture zero corresponds to situation~\ref{situation:1} defined in \Cref{subsec:Definition_of_Modular_Zeros}.}
However, this approach also fails to reproduce the hierarchies of \eqref{eq:toy_mass_matrix}.
In \cite{Liang:2025dkm,Kobayashi:2025ldi} it was argued that non-invertible symmetries can lead to similar results.

A more robust way of obtaining the structure \eqref{eq:toy_mass_matrix} is to use modular zeros. 
If we assign both the quark doublets and $d$-type quarks the representations $(\rep{1''},\rep{1})$ under a modular $T'$ and weight $-\nicefrac{5}{2}$, the mass matrix becomes 
\begin{equation}\label{eq:Weinberg_modular}
 M_\text{modular zero}=\begin{pmatrix}
    0 & \mathcal{Y}_{\rep{1'}}^{(4)}\\ 
    \mathcal{Y}_{\rep{1'}}^{(4)} & \mathcal{Y}_{\rep{1}}^{(4)} 
   \end{pmatrix}\;,
\end{equation}
up to coefficients. 
Note that the mass texture in \Cref{eq:Weinberg_modular} can also be realized within the \ac{BU} framework, where the modular weight of the superpotential vanishes. 
In this case, the modular weight of the quarks needs to be $-2$.

When the modulus field $\ModulusT$ attains an unsuppressed imaginary part, the mass matrix in \Cref{eq:Weinberg_modular} takes a hierarchical structure due to the Fourier expansion of the modular forms. 
This texture is given by   
\begin{equation}\label{eq:modularTextureZero}
 M_\text{modular zero}\xrightarrow{~2\piup\,\im\ModulusT/3>1~} \begin{pmatrix}
    0 & \varepsilon\\ 
    \varepsilon & 1
   \end{pmatrix}\;.
\end{equation}
Here, $\varepsilon\defeq\ee^{\nicefrac{2\pi\ii\ModulusT}{3}}$.
Unlike the texture \eqref{eq:2x2FNmass} obtained from the holomorphic zeros, we see that \eqref{eq:modularTextureZero} has the desired form \eqref{eq:toy_mass_matrix}.

Following Weinberg's argument, we thus conclude that the modular mass matrix can naturally explain the Cabibbo angle and the mass hierarchy of the lighter two generations, $\vartheta_\mathrm{C} \approx \sqrt{m_d/m_s} = \varepsilon = 0.227$.

The above modular texture zero in the two-generation case can be extended to a three-generation scenario.
Note that the ``gaps'' also exist for some non-holomorphic \acp{VVMF} that are relevant for non-supersymmetric theories, such as polyharmonic Maass forms~\cite{Qu:2024rns,Qu:2025ddz}. 
This means that, at least classically, the modular texture zeros can naturally be extended to non-supersymmetric theories. 
We leave a more detailed discussion for future work. 
This includes studying the impact of quantum corrections on the gaps in non-holomorphic scenarios. Moreover, our analysis relied primarily on the superpotential. On the other hand, the K\"ahler potential is not protected by holomorphy, which limits the predictive power of the scheme~\cite{Chen:2019ewa}.

%%%%%%%%%%%%%%%%%%%%%%%%%%%%%%%%%%%%%%%%%%%%%%%%%%%%%%%%%%%%%%%%%%%%%%%%%%
%  Conclusions
%%%%%%%%%%%%%%%%%%%%%%%%%%%%%%%%%%%%%%%%%%%%%%%%%%%%%%%%%%%%%%%%%%%%%%%%%%
\section{Summary}
\label{sec:conclusions}

In this Letter, we have shown that modular symmetries and the properties of modular forms have implications that have not yet been fully appreciated in the physics context. 
There are gaps in the modular form space which give rise to what we call modular zeros. 
This leads to the absence of couplings which is not \emph{due} to symmetries in the usual sense. 
We identified a property of \acp{VVMF}, which we called modular spin, that allows us to understand certain types of stringy zeros. 
As we have shown, modular spin also implies that the R-symmetries in heterotic orbifolds are larger than previously appreciated.

Modular zeros can also be used to address the $\mu$ problem.
In addition, they allow us to address the long-standing question of reproducing Weinberg's famous mass structure for the lighter two down-type quarks. 
As we have shown, if we explain the texture zero through a modular zero, we automatically get the right hierarchies that relate the mass hierarchy to the Cabibbo angle.

Clearly, this is just the beginning of the exploration of these remarkable properties of modular forms in the context of particle physics.

\subsection*{Acknowledgments}

It is a pleasure to thank Mu-Chun Chen and Alexander Stewart for useful discussions.
We further thank V.~Knapp-Perez, Hans Peter Nilles and Sa{\'u}l Ramos-S{\'a}nchez for intense discussions on some results presented in this work, and for sharing their perspective with us.  
This work was supported in part by U.S.\ National Science Foundation grant PHY-2210283.

%%%%%%%%%%%%%%%%%%%%%%%%%%%%%%%%%%%%%%%%%%%%%%%%%%%%%%%%%%%%%%%%%%%%%%%%%%
%  Appendix
%%%%%%%%%%%%%%%%%%%%%%%%%%%%%%%%%%%%%%%%%%%%%%%%%%%%%%%%%%%%%%%%%%%%%%%%%%
\appendix

\section{The Group \texorpdfstring{$T'$}{T'} and Gaps in Its Modular Forms Space}
\label{app:Tprime}

$\Gamma'_3\cong T'$ has 24 elements, and the GAP Id is \texttt{[24,3]}. 
It has the following presentation in terms of the modular group generators $\Generator{S}, \Generator{T}$:
\begin{equation}
 T' = \Braket{\Generator{S}, \Generator{T} ~|~ \Generator{S}^4 = (\Generator{S}\Generator{T})^3 = \Generator{T}^3 = \Generator{S}^2 \Generator{T}\Generator{S}^{-2}\Generator{T}^{-1} = \Id  }\;.
\end{equation}
Its irreducible representations are a triplet $\rep{3}$, three doublets, $\rep{2}$, $\rep{2'}$ and $\rep{2''}$, as well as three 1-dimensional representations $\rep{1}$, $\rep{1'}$ and $\rep{1''}$. The corresponding representation matrices and the \ac{CG} coefficients can be found in \cite{Arriaga-Osante:2025ppz}.

\begin{table}[th!]
\centering
\subcaptionbox{\label{tab:VVMFs}}[0.49\linewidth]{
\begin{tabular}{cc}
\toprule
$k_{\mathcal{Y}}$ & \acp{VVMF} $\mathcal{Y}_{\rep{s}}^{(k_{\mathcal{Y}})}$ \\
\midrule
$1$ & $\mathcal{Y}^{(1)}_{\rep{2''}}$ \\
$2$ & $\mathcal{Y}^{(2)}_{\rep{3}}$ \\
$3$ & $\mathcal{Y}^{(3)}_{\rep{2'}}, ~~\mathcal{Y}^{(3)}_{\rep{2''}}$ \\
$4$ & $\mathcal{Y}^{(4)}_{\rep{1}}, ~~\mathcal{Y}^{(4)}_{\rep{1'}}, ~~\mathcal{Y}^{(4)}_{\rep{3}}$ \\
$5$ & $\mathcal{Y}^{(5)}_{\rep{2}}, ~~\mathcal{Y}^{(5)}_{\rep{2'}}, ~~\mathcal{Y}^{(5)}_{\rep{2''}}$ \\
$6$ & $\mathcal{Y}^{(6)}_{\rep{1}}, ~~\mathcal{Y}^{(6)}_{\rep{3} \textsc{I}}, ~~\mathcal{Y}^{(6)}_{\rep{3} \textsc{II}}$ \\
$7$ & $\mathcal{Y}^{(7)}_{\rep{2}}, ~~\mathcal{Y}^{(7)}_{\rep{2'}}, ~~\mathcal{Y}^{(7)}_{\rep{2''} \textsc{I}}, ~~\mathcal{Y}^{(7)}_{\rep{2''} \textsc{II}}$ \\
$8$ & $\mathcal{Y}^{(8)}_{\rep{1}}, ~~\mathcal{Y}^{(8)}_{\rep{1'}}, ~~\mathcal{Y}^{(8)}_{\rep{1''}}, ~~\mathcal{Y}^{(8)}_{\rep{3} \textsc{I}}, ~~\mathcal{Y}^{(8)}_{\rep{3} \textsc{II}}$ \\
$9$ & $\mathcal{Y}^{(9)}_{\rep{2}}, ~~\mathcal{Y}^{(9)}_{\rep{2'} \textsc{I}}, ~~\mathcal{Y}^{(9)}_{\rep{2'} \textsc{II}}, ~~\mathcal{Y}^{(9)}_{\rep{2''} \textsc{I}}, ~~\mathcal{Y}^{(9)}_{\rep{2''} \textsc{II}}$ \\
\bottomrule
\end{tabular}
}
\hfill
\subcaptionbox{\label{tab:GapsInTprime}}[0.49\linewidth]{%
\begin{tabular}{|c|ccc|ccc|c|}\hline
\diagbox{$k$}{$\rho$} & $\rep{1}$  & $\rep{1'}$  & $\rep{1''}$  &$\rep{2}$ & $\rep{2'}$ & $\rep{2''}$ & $\rep{3}$ \\ \hline
$<0$ &  &  &  & &  &   & \\ \hline
$0$& \FilledEven &  &  & &  &   & \\ \hline
$1$&  & & &  &  & \FilledOdd  &\\ \hline
$2$& & &  &  & & & \FilledEven \\ \hline
$3$& & & &   &  \FilledOdd & \FilledOdd  &\\ \hline
$4$& \FilledEven & \FilledEven &  & & & & \FilledEven \\
\hline
$5$& & & & \FilledOdd   & \FilledOdd  &  \FilledOdd  &\\ \hline
$6$& \FilledEven & &  & & &  & \FilledEven \\ \hline
$7$& &  &  & \FilledOdd  &  \FilledOdd  & \FilledOdd  &  \\ \hline
$8$& \FilledEven & \FilledEven &  \FilledEven  & & & & \FilledEven \\ \hline
$9$& & &  & \FilledOdd  & \FilledOdd  &  \FilledOdd   & \\ 
\hline
\end{tabular}
}
\caption{(a) List of \acp{VVMF} of $\Gamma'_3\cong T'$ up to weight 9. 
The subscripts \textsc{I} and \textsc{II} distinguish different forms with the same $T'$ representation and weight. (b) The distribution of the existence of \acp{VVMF} on $\Gamma'_3\cong T'$ with respect to $(k,\rho)$ up to weight $9$. Empty cells indicate \emph{modular zeros}, i.e. the absence of \acp{VVMF} at the corresponding $(k,\rho)$. The colored cells represent \acp{VVMF} at the corresponding $(k,\rho)$, with orange corresponding to even representations (i.e.\ $\rho(\Generator{S}^2)=1$) and blue to odd ones (i.e.\ $\rho(\Generator{S}^2)=-1$).}
\end{table}
The \acp{VVMF} associated with $\Gamma_3'\cong T'$ can be constructed from tensor products of the doublet \ac{VVMF} $\mathcal{Y}^{(1)}_{\rep{2''}}$ of minimal weight 1. 
Since $\rep{2''}\otimes\rep{2''}=\rep{3}\oplus\rep{1'}$, the antisymmetric contraction $\left(\mathcal{Y}^{(1)}_{\rep{2''}}\otimes\mathcal{Y}^{(1)}_{\rep{2''}}\right)_{\rep{1'}}$ vanishes for identical doublets, while the symmetric contraction $\left(\mathcal{Y}^{(1)}_{\rep{2''}}\otimes\mathcal{Y}^{(1)}_{\rep{2''}}\right)_{\rep{3}}$ leads to a non-vanishing \ac{VVMF} $\mathcal{Y}^{(2)}_{\rep{3}}$ of weight 2.  
A list of \acp{VVMF} for $1\leq k_{\mathcal{Y}} \leq 9$ in $T'$ is shown in \Cref{tab:VVMFs}. \Cref{tab:GapsInTprime} shows the gaps in the modular form space of $\Gamma'_3\cong T'$ more intuitively, that is, the blank cells. 
The colored cells represent non-trivial \acp{VVMF} and the corresponding multiplicities are omitted. 
These results can be obtained either through the algebraic construction of the \acp{VVMF} of $T'$, or directly from the dimension formula of the \acp{VVMF} spaces~\cite[Section~2.2]{Liu:2021gwa}. 
The specific construction of these modular forms of $T'$ can be found in~\cite{Liu:2019khw,Arriaga-Osante:2025ppz}. 
Some typical examples are $\mathcal{Y}^{(4)}_{\rep{1}} = E_4$ and $\mathcal{Y}^{(4)}_{\rep{1'}} = \eta^8$, where $\eta$ and $E_4$ are the Dedekind eta-function and the Eisenstein series, respectively~\cite{DHoker:2022dxx}.
Note that $T'$ is the double cover of $A_4$, or equivalently, $A_4 = T' / \mathds{Z}_2^{ \Generator{S}^2}$; the representation of $A_4$ corresponds to those even representations $\rho(\Generator{S}^2) = \Id$, including $\rep{1}, \rep{1'}, \rep{1''}, \rep{3}$. 
From \Cref{eq:identityonYTrafo}, which may be explained by the modular spin, we know that the \acp{VVMF} of even weight for $T'$ are also \acp{VVMF} of $A_4$; for example, $\mathcal{Y}^{(2)}_{\rep{3}}$ is a modular form triplet of $A_4$.

\section{Holomorphic Zeros}
\label{app:HolomorphicZeros}

Many string models have a \Group{U}{1} factor which appears to be anomalous~\cite{Binetruy:1994ru,Binetruy:1996xk}, and which we will refer to as $\Group{U}{1}_\mathrm{anom}$. 
Note that anomalies of $\Group{U}{1}_\mathrm{anom}$ get cancelled by the \ac{GS} mechanism.

As is well known, the presence of the $\Group{U}{1}_\mathrm{anom}$ symmetry leads to the appearance of an \ac{FI} term at 1-loop~\cite{Fischler:1981zk}. 
The corresponding contribution to the $D$-term potential is given by
\begin{equation}\label{eq:D-term_potential_anomalous_U(1)}
  \mathscr{V}_D 
  \supset
  \frac{1}{2}\left(g_\mathrm{anom}\,\sum_i q_\mathrm{anom}^{(i)}\,\bigl\lvert\Phi^{(i)}\bigr\rvert^2-\xi_\mathrm{FI}\right)^2\;. 
\end{equation}
Here, we sum over all chiral superfields $\Phi^{(i)}$, and $q_\mathrm{anom}^{(i)}$ is the $\Group{U}{1}_\mathrm{anom}$ charge of the $i$\textsuperscript{th} field. 
Further, $g_\mathrm{anom}$ is the coupling strength, and $\xi_\mathrm{FI}$ denotes the \ac{FI} term. In the simplest case, there is a single field, $\Phi_+$, whose \ac{VEV} cancels the \ac{FI} term in \eqref{eq:D-term_potential_anomalous_U(1)}. 

Holomorphy and the sign of the charge of the field(s) that cancel the \ac{FI} term lead to restrictions on the couplings, known as ``holomorphic zeros''~\cite{Leurer:1992wg,Leurer:1993gy}. 
To see this, consider a superpotential term, i.e.\ a monomial of superfields 
\begin{equation}\label{eq:monomial_of_chiral_superfields}
 \mathscr{M}=\prod\limits_i\Phi_i^{n_i}\;.  
\end{equation} 
Since the superpotential $\mathscr{W}$ is holomorphic, it must be a \emph{holomorphic} function of the left-chiral superfields $\Phi_i$. 
For the superpotential to be well-defined for vanishing superfields, the exponents in \eqref{eq:monomial_of_chiral_superfields} have to be nonnegative. 
There are now two cases:
\begin{enumerate}
 \item $q_\mathrm{anom}\bigl(\mathscr{M}\bigr)\le0$, and\label{item:holomorphic_zeros_case_1} 
 \item $q_\mathrm{anom}\bigl(\mathscr{M}\bigr)>0$.\label{item:holomorphic_zeros_case_2}   
\end{enumerate}
In case \ref{item:holomorphic_zeros_case_1}, we can find an $n_+\ge0$ such that a superpotential term $\mathscr{M}\cdot\Phi_+^{n_+}$ is $\Group{U}{1}_\mathrm{anom}$ invariant. 
On the other hand, in case \ref{item:holomorphic_zeros_case_2}, there is no such $n_+$, and therefore the corresponding superpotential term is prohibited. 
The holomorphic zeros caused here by the signs of the charges $q_\mathrm{anom}\bigl(\mathscr{M}\bigr)$ are analogous to the modular zeros that appear in situation 1 discussed in \Cref{subsec:Definition_of_Modular_Zeros}, which are caused by the signs of the modular weights $k_{\mathcal{Y}}$.

Notice that these holomorphic zeros are not expected to be exact. 
In \cite{Araki:2008ek}, it was argued that these zeros can be lifted by multiplying the monomials by coefficients $\ee^{-b\,S}$~\cite{Arkani-Hamed:1998ufq,Dine:2006gm}, where $S$ contains the \ac{GS} axion, and $b\in\mathds{R}$ is appropriately chosen. 
In more detail, as the axion shifts under $\Group{U}{1}_\mathrm{anom}$, combinations of the form 
\begin{equation}
  \mathscr{W}\supset B\,\ee^{-b\,S}\,\mathscr{M}
\end{equation} 
are invariant under $\Group{U}{1}_\mathrm{anom}$ and holomorphic. 
Here, $B$ is a constant. 
However, as these coefficients are exponentially suppressed, they are typically very close to zero. 

Since $\Group{U}{1}_\mathrm{anom}$ is broken and the monomials \eqref{eq:monomial_of_chiral_superfields} are not prohibited by a discrete remnant of $\Group{U}{1}_\mathrm{anom}$, the appearance of these zeros may seem a bit puzzling. 
Since holomorphy plays a role in understanding them, they are often called holomorphic zeros.

%%%%%%%%%%%%%%%%%%%%%%%%%%%%%%%%%%%%%%%%%%%%%%%%%%%%%%%%%%%%%%%%%%%%%%%%%%
%  Bibliography
%%%%%%%%%%%%%%%%%%%%%%%%%%%%%%%%%%%%%%%%%%%%%%%%%%%%%%%%%%%%%%%%%%%%%%%%%%
\bibliography{ModularZeros}
\bibliographystyle{utphys}

\end{document}